\documentclass[12pt]{JHEP3}

\usepackage{amsmath,amsfonts}

\title{Gauduchon-Tod structures,  $Sim$ holonomy \\ and De Sitter supergravity}

\author{Jai Grover \\ DAMTP, Centre for Mathematical Sciences\\
        University of Cambridge\\
        Wilberforce Road, Cambridge, CB3 0WA, UK \\
        \email{jg372@damtp.cam.ac.uk}}
        
\author{Jan B Gutowski \\  Department of Mathematics\\
King's College London, Strand\\
London WC2R 2LS, UK\\
\email{jan.gutowski@kcl.ac.uk}}

\author{Carlos A. R. Herdeiro \\ Departamento de F\'\i sica e Centro de F\'\i sica do Porto
        \\ Faculdade de Ci\^encias da Universidade do Porto\\
        Rua do Campo Alegre, 687, 4169-007 Porto, Portugal \\
        \email{crherdei@fc.up.pt}}
        
\author{ Patrick Meessen and Alberto Palomo-Lozano
\\ Instituto de F\'{\i}sica Te\'orica UAM/CSIC\\
                      Universidad Aut\'onoma de Madrid, C-XVI\\
                      Cantoblanco, E-28670 Madrid, Spain \\
                      \email{patrick.meessen@uam.es, alberto.palomo@uam.es}}

\author{Wafic A. Sabra \\ Centre for Advanced Mathematical Sciences and Physics Department\\
American University of Beirut \\ P.O. Box 11-0236, Riad El-Solh 1107-2020 \\ Beirut,  Lebanon \\
\email{ws00@aub.edu.lb}}

\abstract{Solutions of five-dimensional De Sitter supergravity admitting Killing spinors are considered, using spinorial geometry techniques. It is shown that the ``null'' solutions are defined in terms of a one parameter family of 3-dimensional constrained Einstein-Weyl spaces called Gauduchon-Tod structures. 
They admit a geodesic, expansion-free, twist-free and shear-free null vector field and therefore are a particular type of Kundt geometry. When the Gauduchon-Tod structure reduces to the 3-sphere, the null vector becomes recurrent, and therefore the holonomy is contained in $Sim(3)$, the maximal proper subgroup of the Lorentz group $SO(4,1)$. For these geometries, all scalar invariants built from the curvature are constant. Explicit examples are discussed.}


\begin{document}

\section{Introduction}
Genuine supergravity theories can have a vanishing or a negative cosmological constant, but not a positive one \cite{Pilch:1984aw,Lukierski:1984it}.  In the latter case one may, however, introduce the concept of ``fake" supergravity as a solution generating technique \cite{Skenderis:2006jq}.  Recently \cite{guthktd5} we have initiated the programme of determining all solutions admitting (pseudo-)Killing spinors in De Sitter ``supergravity" theories. In \cite{guthktd5} the ``timelike" case of minimal De Sitter supergravity in five dimensions was analysed. The resulting geometries are defined in terms of a four dimensional base space which is a hyper-K\"ahler manifold with torsion (HKT) and a set of constraint equations. Together with the minimal ungauged (i.e Minkowski) and gauged (i.e Anti-De-Sitter) supergravity theories in five dimensions, this result established that all (pseudo-)supersymmetric geometries of five dimensional minimal supergravities are defined in terms of four dimensional complex geometries, namely HKT, hyper-K\"ahler and K\"ahler manifolds \cite{Grover:2009ze}. 

In this paper we shall analyse the null case of minimal De Sitter supergravity in five dimensions, using spinorial geometry techniques \cite{papadopd11,papadopiib,d11preon,iibpreon,papadtype1,gutpapd4,roestd4}. Our main result is the following: all solutions of the minimal five dimensional De Sitter supergravity theory admitting (pseudo)-Killing spinors from which a null vector field can be constructed fall into the following family of backgrounds:
\begin{equation}
ds^2 = 2 du \bigg( dv + \big(H-{\chi^2 \over 8} v^2\big) du + \chi v {{\cal{B}}}+ \phi 
\bigg) - ds_{GT}^2  \ , \qquad 
F = {\chi \over 4} du \wedge dv+d{{\cal{B}}} \  , \label{solution}
\end{equation}
where $\chi^2/2$ is the cosmological constant, $GT$ is $u$-dependent Gauduchon-Tod space \cite{tod}, $H$, ${{\cal{B}}}$ and $\phi$ are, respectively, a function and two 1-forms on $GT$ which may also depend on $u$ (but not on $v$). The constraints on $GT$, ${{\cal{B}}}$, $\phi$ and $H$ are summarised in section \ref{summary}.

Gauduchon-Tod spaces were initially discussed in the context of hyper-hermitian spaces admitting a tri-holomorphic Killing vector field. They are special types of Einstein-Weyl 3-spaces, obeying constraints. Curiously, these spaces play also a role in the timelike class of solutions in both $D=5$ and in $D=4$. Since GT spaces define a four dimensional HKT geometry, they were  used in \cite{guthktd5} to construct examples of timelike solutions of the $D=5$ minimal De Sitter supergravity theory for which the base space is not conformally hyper-K\"ahler. In the $D=4$ minimal De Sitter theory, the timelike solutions are defined by a base space which is GT \cite{Meessen:2009ma}. But whereas the Ricci curvature of the Weyl connection is always non-flat in the solutions we describe in this paper, the $D=4$ timelike solutions allow flat GT spaces.

As for the null supersymmetric solutions of the minimal five dimensional ungauged \cite{Gauntlett:2002nw} and gauged \cite{Gauntlett:2003fk} theories,
the family of backgrounds \eqref{solution} admits a geodesic, expansion-free, twist-free and shear-free null vector field $N$. In four dimensional General Relativity, such geometries are dubbed \textit{Kundt metrics} \cite{Kundt:1961}. In higher dimensions, these geometries have been considered in  \cite{Coley:2009ut,Podolsky:2008ec,Brannlund:2008zf}. But $N$ has distinct properties in the De Sitter theory, as compared with the Minkowski or AdS theories. In the Minkowski and AdS case, the null vector is always Killing; and for some special cases  it becomes covariantly constant. Then the Kundt geometries become plane-fronted waves with parallel rays (\textit{pp-waves}). This is not the case for the De Sitter theory. For  the special case with ${{\cal{B}}}=0$, however, the null vector acquires an interesting property: it becomes  \textit{recurrent}, that is, it obeys 
\begin{equation}
\nabla_\mu N^\nu = C_\mu N^\nu \ , \label{recurrence0}
\end{equation}
for some non-trivial, recurrence one form $C_{\mu}$. This means that the geometries \eqref{solution} have special holonomy $Sim(3)$, which is the maximal proper subgroup of the Lorentz group $SO(4,1)$. 

The four parameter Similitude group, $Sim(2)$, became a focus of recent interest due to the proposal, by Cohen and Glashow, of \textit{Very Special Relativity} (VSR) \cite{Cohen:2006ky}. These authors investigated if the exact symmetry group of nature could be isomorphic to a proper subgroup of the Poincar\'e group rather than the Poincar\'e group itself. The proper subgroup they considered was obtained by adjoining the maximal proper subgroup of the Lorentz group, $Sim(2)$, with spacetime translations. The theory based on this symmetry group, VSR, actually implies Special Relativity if a discrete symmetry, namely CP, is also added. But since the latter is broken in nature, VSR is distinct from Special Relativity, albeit many sensitive searches for departures from Lorentz invariance will fail if VSR is the true symmetry of nature. In a subsequent development \cite{Gibbons:2007iu}, it was shown that \textit{General Very Special Relativity}, i.e. a theory based on a symmetry group obtained by a continuous deformation of the Inhomogeneous $Sim(2)$ group, $ISim(2)$, is a \textit{Finslerian} geometry, since the invariant line element, which is a homogeneous function of degree one in displacements, is not quadratic and it is anisotropic. 

Perhaps partly motivated by the Cohen and Glashow proposal, studies of $d$ dimensional Lorentzian geometries with $Sim(d-2)$  holonomy have been carried out recently \cite{Gibbons:2007zu}. The resulting geometries have interesting properties, such as the possibility of vanishing quantum corrections \cite{Coley:2008th}. Possible connections to supersymmetry have also been hinted at \cite{Brannlund:2008zf}. Here, we show how these geometries indeed emerge in an explicit supersymmetry computation, a fact recently unveiled in a four dimensional example \cite{Meessen:2009ma} (see also \cite{Gutowski:2009vb}).

This paper is organised as follows. In section 2 we describe the theory to be considered as well as some generalities of the spinorial geometry technique that shall be used. Section 3 gives the details of the calculations leading to the result \eqref{solution}. Properties of the resulting geometries are described in section 4, wherein a brief comparison with the null supersymmetric solutions of the minimal ungagued and gauged supergravities in five dimensions is also performed. We then focus on the special case with ${{\cal{B}}}=0$ in \eqref{solution}, which is the most general solution for which the null vector field $N$ is recurrent, and discuss special properties of the curvature for this solutions. Examples with ${{\cal{B}}}=0$ and ${{\cal{B}}}\neq 0$ are presented. Final remarks are given in section 5. Some other technical details of the computation are described in two appendices. A third appendix presents an introduction to Gauduchon-Tod spaces.

\section{Minimal $D=5$ De Sitter Supergravity}

We begin with a brief review of $\mathcal{N}=2$, $D=5$ minimal De Sitter
supergravity. The fake gravitino Killing spinor equation for this
theory is\footnote{In this paper we shall use a mostly minus signature.}
\begin{equation}\label{kse}
\left( \partial _{\mu }+{1\over4}\omega _{\mu }{}^{\rho \sigma
}\Gamma _{\rho \sigma }-{\chi\over2}A_{\mu
}+{\chi\over4\sqrt{3}}\Gamma _{\mu }-{\sqrt{3}\over2}F_{\mu
}{}^{\rho }\Gamma _{\rho }+{1\over4\sqrt{3}}\Gamma _{\mu }F^{\rho
\sigma }\Gamma _{\rho \sigma }\right) \epsilon =0\ ,
\end{equation}
where $\epsilon $ is a Dirac spinor. Here $F = dA$ is the gauge
field strength and $\chi$ is a non-zero real constant. The metric
has vielbein $\mathbf{e}^{+},\mathbf{e}^{-},\mathbf{e}^{1},
\mathbf{e}^{{\bar{1}}},\mathbf{e}^{2}$, where $\mathbf{e}^{\pm },
\mathbf{e}^{2}$ are real, and
$\mathbf{e}^{1},\mathbf{e}^{{\bar{1}}}$ are complex conjugate, and
\begin{equation}
ds^{2}=2\mathbf{e}^{+}\mathbf{e}^{-}-2\mathbf{e}^{1}\mathbf{e}^{\bar{1}}
-(\mathbf{e}^{2})^{2}\ . 
\end{equation}

The Einstein and gauge field equations are expressed as

\begin{equation}
 R_{\mu\nu} + 2F_{\mu\sigma}F_{\nu}{}^{\sigma} -
{g_{\mu\nu}\over3}(F^2 - \chi^2) = 0 \ ,  
\end{equation}
and
\begin{equation}
d*F + {2\over\sqrt{3}}F\wedge F = 0 \ ,  
\end{equation}
respectively, where $F^2 = F_{\rho\sigma}F^{\rho\sigma}$. We should
note that, unlike the timelike case, for these solutions one
component of the Einstein equations must be imposed in addition to
the Killing spinor and gauge equations (see \cite{Gauntlett:2003fk,guthktd5} for a more detailed discussion on this point).

For De Sitter supergravity in five-dimensions, one takes the space
of Dirac spinors to be the space of complexified forms on ${\mathbb{R}}^2$,
which are spanned over ${\mathbb{C}}$ by $\{ 1, e_1, e_2, e_{12} \}$ where
$e_{12}=e_1 \wedge e_2$. The action of complexified
$\gamma$-matrices on these spinors is given by

\begin{eqnarray}
\gamma_j = i (e_j \wedge + i_{e_j})\ ,  
\qquad 
\gamma_{j+2} = -e_j \wedge +i_{e_j} \ ,  
\end{eqnarray}
for $j=1,2$. $\gamma_0$ is defined by
\begin{equation}
\gamma_0 = \gamma_{1234} \ ,  
\end{equation}
and satisfies
\begin{equation}
\gamma_0 1 = 1 , \quad \gamma_0 e_{12} = e_{12} , \quad \gamma_0 e_j
= -e_j \ ,\qquad  j=1,2 \ .  
\end{equation}

In what follows we will restrict our attention to the constraints
obtained from the Killing spinor equation ({\ref{kse}}) in the null
case, i.e. when the vector field constructed from the Killing spinor is
null. It will then be useful to adopt a null basis in the
$\gamma$-matrices

\begin{eqnarray}
\Gamma _{\pm } &=&{\frac{1}{\sqrt{2}}}(\gamma _{0}\mp \gamma _{3})\ ,  \nonumber \\
\Gamma _{1} &=&{\frac{1}{\sqrt{2}}}(\gamma _{2}-i\gamma
_{4})=\sqrt{2}
ie_{2}\wedge \ ,  \nonumber \\
\Gamma _{\bar{1}} &=&{\frac{1}{\sqrt{2}}}(\gamma _{2}+i\gamma
_{4})=\sqrt{2}
ii_{e_{2}}\ ,  \nonumber \\
\Gamma _{2} &=&\gamma _{1}\ .  
\end{eqnarray}

Finally, as in \cite{gutnulld5}, we can put a generic null Killing
spinor into a simple canonical form
\begin{equation}
\epsilon =1+e_{1} \ ,  
\end{equation}
by making use of $Spin(4,1)$ gauge transformations. The resulting
equations, obtained by evaluating the Killing spinor equation on
$\epsilon$, are listed in Appendix A.
We remark also that if $\epsilon=1+e_1$ satisfies the Killing spinor 
equations, then so does the spinor $e_2-e_{12}$. This can be seen 
by noting that the operator $C$ defined via
\begin{equation}
C 1 = -e_{12}, \quad C e_{12} = 1, \quad C e_1 = e_2, \quad C e_2 = - e_1
\end{equation}
satisfies
\begin{equation}
C * \gamma_\mu = \gamma_\mu C * \ .
\end{equation}
It therefore follows that if $\epsilon$ satisfies ({\ref{kse}}) then so does $C* \epsilon$.
Hence the solutions under consideration here preserve at least half of the (pseudo)-supersymmetry.\footnote{Since the action of $C \ *$ does not depend on the timelike or null class, we conclude that,  
for the timelike solutions obtained in \cite{guthktd5}, there is again at least one half of (pseudo)-supersymmetry preserved, the Killing spinors in this case being, at least, $1$ and $e_{12}$.}

\section{Analysis of the Constraints}

An analysis of the equations presented in Appendix A, yields the
following relations between the gauge potential and the spin connection
\begin{equation}
A_{+} =-\frac{1}{\chi }\omega _{+,+-} \ , \qquad A_{-} =-\frac{1}{{\chi }}\omega _{-,+-} \ ,\end{equation}
\begin{equation}
A_{1} =-\frac{1}{\chi }\left(\omega _{1,+-}+ \omega _{-,+1}\right)
\ , \qquad  A_{2} =-\frac{1}{{\chi }}\left( \omega _{2,+-}+\omega
_{-,+2}\right) \ ;
\end{equation}
between the field strength and the spin connection
\begin{equation}
\text{\ \ }F_{1\bar{1}} =-i\sqrt{3}\omega _{-,+2}\ , \qquad 
F_{12} = -\frac{\sqrt{3}}{2}i\omega _{2,12} \ , \qquad
F_{+-} = -\frac{1}{4}{\chi }\ , \end{equation}
\begin{equation}
F_{-1} = \frac{i}{\sqrt{3}}\omega _{-,12} \ , \qquad
\text{\ }F_{+2} =F_{+1}=0, \ , \qquad
F_{-2} =-\frac{i}{\sqrt{3}}\omega _{-,1\bar{1}}\ ;
\end{equation}
and the following constraints on the spin connection
\begin{equation}
\omega _{2,+\bar{1}} =\omega _{2,+2}=\omega _{+,+2} =\omega _{+,+\bar{1}}=\omega _{+,\bar{1}2}\ =
\omega _{+,1\bar{1}}= \omega _{\bar{1},\bar{1}2} =\omega _{\bar{1},+\bar{1}}=\omega
_{1,+2} =\omega _{1,+\bar{1}}=0 \ ,
\end{equation}
\begin{equation}
\label{geom1}
\omega _{2,1\bar{1}} ={\frac{\sqrt{3}i\chi }{4}}, \qquad
\omega _{2,\bar{1}2} =-2\omega _{-,+\bar{1}}= \omega
_{_{\bar{1}},_{\bar{1}1}} \ ,
\end{equation}
and 
\begin{equation} 
-2\omega _{-,+2}+\omega _{1,\bar{1}2}-\frac{\sqrt{3}}{4}i{\chi } =0\ . \label{geom2} \end{equation}
Thus, the gauge field one-form is given by
\begin{eqnarray} \chi A &=&\big( -\omega _{+,+-}\big) \mathbf{e}^{+}-\omega
_{-,+-} \mathbf{e}^{-} - \big( \omega _{1,+-}+ \omega _{-,+1}\big)
\mathbf{e}^{1} \nonumber \\ &-& \big( \omega _{\bar{1},+-}+ \omega
_{-,+\bar{1}}\big) \mathbf{e}^{\bar{1}}-\big( \omega _{2,+-}+\omega
_{-,+2}\big) \mathbf{e}^{2} \ ,   \end{eqnarray}
and the field strength 2-form is given by
\begin{eqnarray}\label{fgeom}
F &=&  -\sqrt{3}i\omega _{-,+2}\mathbf{e}^{1}\wedge
\mathbf{e}^{\bar{1}} -
{\sqrt{3}i\over2}\omega_{2,12}\mathbf{e}^1\wedge \mathbf{e}^2
+\frac{\sqrt{3}}{2}i\omega _{2,\bar{1}2}\mathbf{e}^{\bar{1}}\wedge
\mathbf{e}^{2}  
 -\frac{\chi}{4}\text{ }\mathbf{e}^{+}\wedge
\mathbf{e}^{-} \nonumber \\
&+& {i\over\sqrt{3}}\omega_{-,12}\mathbf{e}^-\wedge \mathbf{e}^1-
\frac{i}{\sqrt{3}}\omega _{-,\bar{1}2}\mathbf{e}^{-}\wedge \mathbf{e}^{\bar{1}}-
\frac{i}{\sqrt{3}}\omega _{-,1\bar{1}}\mathbf{e}^{-}\wedge
\mathbf{e}^{2} \ .
\end{eqnarray}
These constraints are sufficient to imply that
\begin{eqnarray}
(\mathcal{L}_{N}{\mathbf{e}}^-)_m &=& (\mathcal{L}_{N}{\mathbf{e}}^{\alpha})_m =
(\mathcal{L}_{N}{\mathbf{e}}^2)_m = 0 \ ,\nonumber \\ (\mathcal{L}_{N}{\mathbf{e}}^-)_- &=&
-{1\over2}\omega_{+,+-} \ , (\mathcal{L}_{N}{\mathbf{e}}^{\alpha})_- =
-{1\over2}(\omega_{-,+\bar{\alpha}} - \omega_{+,-\bar{\alpha}}) \ ,
\nonumber \\ (\mathcal{L}_{N}{\mathbf{e}}^2)_- &=& -{1\over2}(\omega_{-,+2}
-\omega_{+,-2})\ ,  
\end{eqnarray}
for $m = 1, \bar{1}, 2$, and where we have introduced a coordinate
$v$ such that 
\begin{equation}
N = {\mathbf{e}}_+ = {\partial\over\partial v} \ . \end{equation}
The non-zero
components of these Lie derivatives can be eliminated by making use
of the residual gauge freedom; those transformations which leave
$\epsilon = 1 + e_1$ invariant. The details are presented in Appendix B.
We therefore set, without loss of generality, $A_+=0$, and
\begin{eqnarray}\label{lieminus}
\mathcal{L}_{N}{\mathbf{e}}^- = 0 \ ,\qquad \mathcal{L}_{N}{\mathbf{e}}^{\alpha} = 0 \ ,
\end{eqnarray}
with
\begin{eqnarray}
\omega_{+,-1} = \omega_{-,+1} \ ,\qquad \omega_{+,-2} =
\omega_{-,+2} \ . 
\end{eqnarray}
Collecting these results we can write the exterior derivatives of
the vielbein as
\begin{eqnarray} d{\mathbf{e}}^+ &=& {\mathbf{e}}^+ \wedge \chi A - \omega_{-,-1}{\mathbf{e}}^- \wedge
{\mathbf{e}}^1 - \omega_{-,-\bar{1}}{\mathbf{e}}^- \wedge {\mathbf{e}}^{\bar{1}}
-\omega_{-,-2}{\mathbf{e}}^-\wedge{\mathbf{e}}^2  \nonumber \\ &-& (\omega_{1,-\bar{1}} 
  -\omega_{\bar{1},-1}) {\mathbf{e}}^1 \wedge {\mathbf{e}}^{\bar{1}} - (\omega_{1,-2}
- \omega_{2,-1}){\mathbf{e}}^1 \wedge {\mathbf{e}}^2 -(\omega_{\bar{1},-2} -
\omega_{2,-\bar{1}}){\mathbf{e}}^{\bar{1}}\wedge {\mathbf{e}}^2 \ , \nonumber \\  \end{eqnarray}
\begin{eqnarray}
 d{\mathbf{e}}^- =
\left[(\omega_{-,+1} - \omega_{1,+-}){\mathbf{e}}^1 + (\omega_{-,+\bar{1}} -
\omega_{\bar{1},+-}){\mathbf{e}}^{\bar{1}} + (\omega_{-,+2} -
\omega_{2,+-}){\mathbf{e}}^2 \right] \wedge {\mathbf{e}}^- \ , 
\end{eqnarray}
\begin{eqnarray}
d{\mathbf{e}}^{1} = \left[(\omega_{-,1\bar{1}} - \omega_{1,-\bar{1}}){\mathbf{e}}^1 -
\omega_{\bar{1},-\bar{1}}{\mathbf{e}}^{\bar{1}} - (\omega_{-,\bar{1}2} +
\omega_{2,-\bar{1}}){\mathbf{e}}^2 \right]\wedge {\mathbf{e}}^- \nonumber \\ +
\omega_{\bar{1},1\bar{1}}{\mathbf{e}}^1 \wedge {\mathbf{e}}^{\bar{1}} +
(\omega_{1,\bar{1}2} + \omega_{2,1\bar{1}}){\mathbf{e}}^1\wedge {\mathbf{e}}^2 \ , 
\end{eqnarray}
\begin{eqnarray}
 d{\mathbf{e}}^{\bar{1}} =
-\left[\omega_{1,-1}{\mathbf{e}}^{1} + (\omega_{-,1\bar{1}} +
\omega_{\bar{1},-1}){\mathbf{e}}^{\bar{1}} +(\omega_{-,12} +
\omega_{2,-1}){\mathbf{e}}^2 \right]\wedge {\mathbf{e}}^- \nonumber \\ +
\omega_{1,1\bar{1}}{\mathbf{e}}^1 \wedge {\mathbf{e}}^{\bar{1}} + (\omega_{\bar{1},12}
- \omega_{2,1\bar{1}}){\mathbf{e}}^{\bar{1}}\wedge {\mathbf{e}}^2 \ , 
\end{eqnarray}
\begin{eqnarray}
d{\mathbf{e}}^2 = \left[(\omega_{-,12} - \omega_{1,-2}){\mathbf{e}}^1 +
(\omega_{-,\bar{1}2} - \omega_{\bar{1},-2}){\mathbf{e}}^{\bar{1}} -
\omega_{2,-2}{\mathbf{e}}^2 \right] \wedge {\mathbf{e}}^- \nonumber \\ -(\omega_{1,\bar{1}2} -
\omega_{\bar{1},12}){\mathbf{e}}^1 \wedge {\mathbf{e}}^{\bar{1}} +
\omega_{2,12}{\mathbf{e}}^1\wedge {\mathbf{e}}^2 +
\omega_{2,\bar{1}2}{\mathbf{e}}^{\bar{1}}\wedge {\mathbf{e}}^2 \ . 
\end{eqnarray}

As ${\mathbf{e}}^{-}$ is hypersurface orthogonal, it is natural to define a coordinate $u$ such that
\begin{equation}
{\mathbf{e}}^- = fdu \ , 
\end{equation}
where $f \in {\mathbb{R}}$ is $v$-independent. We can set $f=1$ by making
a combined ${\mathbb{R}} \times Spin(4,1)$ transformation of the form $e^{-h}
e^{h \Gamma_{+-}}$ for $h \in {\mathbb{R}}$, with $\partial_+ h = 0$. This transformation
leaves $1+e_1$ invariant, and also preserves the gauge $A_+ = 0$. With this
choice ${\mathbf{e}}^-$ is closed, and therefore
\begin{eqnarray}
\omega_{-,+2} = \omega_{2,+-} \ , \qquad \omega_{-,+\alpha} =
\omega_{\alpha,+-} \ , 
\end{eqnarray}
for $\alpha = 1,\bar{1}$. Further progress can be made by examining
the consistency conditions, $F=dA$; from the $(dA)_{+-}$ component
we find
\begin{equation}
\partial_+ A_- = -{\chi\over4} \ , 
\end{equation}
and
\begin{equation}
\partial_+ A_1 = \partial_+ A_{\bar{1}} = \partial_+ A_2 = 0 \ , 
\end{equation}
from the $(dA)_{+1}$, $(dA)_{+\bar{1}}$, and $(dA)_{+2}$ components
respectively.

Next, notice that
\begin{eqnarray}
\mathcal{L}_{N} A = -{\chi\over4}{\mathbf{e}}^- \ ,\qquad \mathcal{L}_{N}{\mathbf{e}}^+
= \chi A \ ,
\end{eqnarray}
together with ({\ref{lieminus}}) imply
\begin{equation}
\mathcal{L}_{N}\mathcal{L}_{N}{\mathbf{e}}^+ = -{\chi^2\over4}{\mathbf{e}}^- \ , \qquad 
\mathcal{L}_{N}\mathcal{L}_{N}A = 0 \ . 
\end{equation}
We can make explicit the $v$-dependance of A and ${\mathbf{e}}^+$ using the
relations above
\begin{equation}
 A = -{\chi\over4}vdu + {{\cal{B}}} \ , 
\end{equation}
\begin{equation}
\label{eplus}
{\mathbf{e}}^+ = dv - {\chi^2\over8}v^2du  + \chi{{\cal{B}}} v + \alpha \ ,
\end{equation}
where ${{\cal{B}}}, \alpha$ are $v$-independent. $F$ then takes the form
\begin{equation}\label{fcoord}
F = -{\chi\over4}dv \wedge du + d{{\cal{B}}} \ ,
\end{equation}
or equivalently
\begin{equation}
F = -{\chi\over4}{\mathbf{e}}^+ \wedge {\mathbf{e}}^-  + {\chi\over4}(\chi v{{\cal{B}}}
+ \alpha)\wedge {\mathbf{e}}^- + d{{\cal{B}}} \ .  \label{effe}
\end{equation}
Having introduced the co-ordinates $u, v$, three remaining real co-ordinates $x^m$ ($m=1,2,3$) can
be introduced such that 
\begin{equation}
{{\bf{e}}}^1 = {{{\bf{e}}}^1}_m dx^m, \qquad {{\bf{e}}}^2 = {{{\bf{e}}}^2}_m dx^m \ .
\end{equation}
Here we have
removed any $du$ terms from ${{\bf{e}}}^1, {{\bf{e}}}^2$ by making use of a gauge transformation of the
form ({\ref{gaugeframe}}). We also
eliminate ${{\cal{B}}}_u$ with a shift in $v$ and a subsequent
redefinition of $\alpha$.

Next consider the constraints ({\ref{geom1}}) and ({\ref{geom2}}); these are equivalent to
\begin{eqnarray}
{{\tilde{d}}} {{\bf{e}}}^2 &=& -{\sqrt{3} i \chi \over 2} {{\bf{e}}}^1 \wedge {{\bf{e}}}^{\bar{1}} - \chi {{\bf{e}}}^2 \wedge {{\cal{B}}} \ ,
\nonumber \\
{{\tilde{d}}} {{\bf{e}}}^1 &=& -{\sqrt{3} i \chi \over 2} {{\bf{e}}}^2 \wedge {{\bf{e}}}^1 - \chi {{\bf{e}}}^1 \wedge {{\cal{B}}} \ ,
\end{eqnarray}
where ${{\tilde{d}}}$ denotes the restriction of the exterior derivative to hypersurfaces of constant $v, u$.
This implies that the 1-parameter family of 3-manifolds $GT$ with metric
\begin{equation}
ds^2_{GT} = ({{\bf{e}}}^2)^2+2 {{\bf{e}}}^1 {{\bf{e}}}^{\bar{1}} \ ,
\end{equation}
admits a real basis ${\bf{E}}^i$ for $i=1,2,3$ such that
\begin{equation}
\label{gtodstruc}
{{\tilde{d}}} {\bf{E}}^i = -{\sqrt{3} \chi \over 2} \star_3 {\bf{E}}^i + \chi {{\cal{B}}} \wedge {\bf{E}}^i \ ,
\end{equation}
where $\star_3$ denotes the Hodge dual on $GT$, with volume form $\epsilon_3 = i {{\bf{e}}}^{1 \bar{1} 2}$.
It follows that $GT$ admits a Gauduchon-Tod structure \cite{tod} (see appendix C for a discussion of these structures).  Note in particular that
({\ref{gtodstruc}}) implies 
\begin{equation}
\label{intc1}
{{\tilde{d}}} {{\cal{B}}} = {\sqrt{3} \chi \over 2} \star_3 {{\cal{B}}} \ ,
\end{equation}
from which we obtain  
\begin{equation}
{{\tilde{d}}} \star_3 {{\cal{B}}}=0 \ .
\end{equation}

To proceed further, compare the expression ({\ref{fcoord}}) for $F$ to
({\ref{fgeom}}), to obtain
\begin{equation}
\label{y1}
Y_1 = {-i\over\sqrt{3}}\omega_{-,12} \ , \qquad Y_2 =
{i\over\sqrt{3}}\omega_{-,1\bar{1}} \ , 
\end{equation}
where
\begin{equation}
\label{y2}
Y_m = {\chi^2v\over4}{{\cal{B}}}_m + {\chi\over4}\alpha_m +
(d{{\cal{B}}})_{m-} \ ,
\end{equation}
for $m = 1, \bar{1}, 2$.

In order to investigate the constraints ({\ref{y1}}), it will be useful to write
\begin{equation}
\alpha = \phi + H du \ ,
\end{equation}
where $\phi= \phi_m dx^m$, and also denote the Lie derivative with
respect to $\partial / \partial u$ as ${{\dot{{{\cal{B}}}}}} = {{\cal{L}}}_{\partial / \partial u} {{\cal{B}}}$.
Then ({\ref{y1}}) is equivalent to
\begin{equation}
\label{conxx1}
{\chi \over 4} \phi - {{\dot{{{\cal{B}}}}}} -{1 \over 2 \sqrt{3}} \star_3 ({{\tilde{d}}} \phi + \chi {{\cal{B}}} \wedge \phi - {\bf{E}}^i \wedge {\dot{\bf{E}}}^i) =0 \ .
\end{equation}
It is straightforward, but tedious, to show that these constraints, together with their associated integrability conditions, 
are sufficient to imply that the gauge field equations hold with no further constraint.
Finally, we consider the Einstein equations. Pseudo-supersymmetry implies that all 
components of the Einstein equations hold automatically, with the exception of the $--$ component,
which must be computed explicitly. From this component, we find the following condition on the function $H$:

\begin{eqnarray}
\label{eineq1}
\Box_3 H + \chi {{\cal{B}}} \cdot {{\tilde{d}}} H = {\tilde{\nabla}}^i {\dot{\phi}}_i + (\ddot{\bf{E}}^i)_i
+ \chi \phi \cdot {{\dot{{{\cal{B}}}}}} -4 {{\dot{{{\cal{B}}}}}}^2 -2 \sqrt{3} \star_3 ({\chi \over 4} \phi - {{\dot{{{\cal{B}}}}}})_{ij}\ (\dot{\bf{E}}^i)_j \ , 
\nonumber \\
\end{eqnarray}
where $\Box_3$ denotes the Laplacian on $GT$.

\subsection{Summary}
\label{summary}
To summarise, all null solutions of minimal five dimensional De Sitter supergravity are constructed as follows:
\begin{enumerate}
\item[i)] Choose a Gauduchon-Tod space $GT$, $
ds^2_{GT}=\delta_{ij}{\bf{E}}^i{\bf{E}}^j$, where the frames obey \eqref{gtodstruc}. ${\bf{E}}^i$ and ${{\cal{B}}}$, in general, depend on $u$.
\item[ii)] Choose a 1-form on $GT$, $\phi$, possibly $u$ dependent, obeying \eqref{conxx1}.
\item[iii)]  Choose a function on $GT$, $H$, possibly $u$ dependent, obeying \eqref{eineq1}.
\item[iv)] The solution is then given by  \eqref{solution}. Note that ${\bf E}^i$, ${{\cal{B}}}$, $\phi$ and $H$ do not depend on the spacetime coordinate $v$.
\end{enumerate}

\section{Properties of the solution and special cases}
The general solution \eqref{solution} is a Kundt geometry. To see this consider the null vector field
$N={\partial / \partial v}$. 
It is straightforward to check that the null congruence with tangent vector $N$ is geodesic ($N^\mu \nabla_\mu N^\nu =0$), hypersurface orthogonal ($N\wedge dN=0$, where $N$ is the 1-form dual to the null vector field), expansion free ($\nabla_\mu N^\mu=0$) and
shear free (since it is expansion free and $\nabla_{(\mu} N_{\nu)}\nabla^\mu N^\nu=0$). It follows that the geometry is of (higher dimensional generalisation of) Kundt type (see \cite{exact} for a thorough discussion of the four dimensional Kundt geometries). It is a special case of the general form presented in \cite{Podolsky:2008ec} for higher dimensional Kundt geometries (see eq. (77) therein).

A distinct feature of the general solution \eqref{solution} when compared to the other null solutions of minimal supergravity theories in $D=5$ is that $N$ is not a Killing vector field. In both the ungauged \cite{Gauntlett:2002nw} and gauged \cite{Gauntlett:2003fk} minimal five dimensional supergravity theories, the general null solution can be written as 
\begin{equation} ds^2=\frac{2du}{H}\left(dv+\left[\mathcal{F}-H^3{\vec b}\cdot {\vec b}\right]\frac{du}{2}-H^3{\vec a}\cdot d{\vec x}\right)-H^2\gamma_{ij}dx^idx^j \ . \end{equation}
In the ungauged case, 
\begin{equation} \vec{b}=\vec{a} \ , \qquad \gamma_{ij}={\rm diag}(1,1,1) \ ; \end{equation}
in the gauged case,
\begin{equation} \vec{b}=\left(a^1,\frac{a^2}{S},\frac{a^3}{S}\right) \ , \qquad \gamma_{ij}={\rm diag}(1,S^2,S^2) \ ; \end{equation}
in both cases the metric functions $H$, $\mathcal{F}$ and vector ${\vec a}$ (with components $a^i$) depend on $(u,{\vec x})$, but not on $v$. The same is true for the function $S$, which appears in the gauged case. In either case a (different) set of constraints has to be obeyed in order to have a susy solution of the theory. 

The null vector field $N=\partial/\partial v$ is therefore Killing and obeys
\begin{equation} \nabla_{\mu}N_{\nu}=N_{[\mu}\partial_{\nu]}\ln H \ . \end{equation}
Generically, the solutions may be characterised as plane-fronted waves, i.e they possess a geodesic, expansion-free, twist-free and shear-free null vector field $N$. If $H$ depends solely on $u$, $N$ becomes covariantly constant and the solutions become plane-fronted waves with parallel rays (pp-waves). 

Another (related) distinction between the null solutions presented here and those of the ungauged and gauged theories is that, for the latter, the null vector is never recurrent. The null vector field $N$ is recurrent if \eqref{recurrence0} holds. But the Killing character of $N$ prevents this possibility. In the De Sitter case, however, the possibility of recurrence arises. A simple calculation shows that this requires $g_{ui,v}=0$. Therefore $N$ is recurrent iff ${{\cal{B}}}=0$. 

A different way of reaching the same conclusion comes about by realising that the 
vector-field $N$ can be constructed as
the vector-bilinear of the Killing spinors (see e.g. ref.~\cite{Gauntlett:2002nw}). This 
identification allows us to derive
the constraint
\begin{equation}\label{eq:trut}
Ê \nabla_{\mu}N_{\nu}\; =\; \chi\ A_{\mu}\ N_{\nu} \; +\; \textstyle{1\over\sqrt{3}}\ 
\left(\imath_{N}\star F\right)_{\mu\nu} \; ,
\end{equation}
which implies that $N$ is recurrent iff $F$ satisfies the radiation condition $N\wedge 
F=0$. Combining this with eqs. (\ref{effe}) and
(\ref{intc1}), then implies that the holonomy of the solution is $Sim$ whenever 
$\mathcal{B}=0$.
\par
The general solution with $\mathcal{B}\neq 0$ can also be given a $Sim$-holonomy 
structure: rewrite eq.~(\ref{eq:trut}) by introducing a torsionful connection
$\mathfrak{D}$ such that
\begin{equation}
Ê \mathfrak{D}_{\mu}N_{\nu} \; \equiv\; \nabla_{\mu}N_{\nu} \ -\ 
S_{\mu\nu}{}^{\sigma}N_{\sigma} \; =\; \chi\ A_{\mu}N_{\nu}
Ê\hspace{.3cm}\mbox{with}\hspace{.3cm} \sqrt{3}S_{\mu\nu\sigma}= \left(\star 
F\right)_{\mu\nu\sigma} \; ,
\end{equation}
so that $N$ is recurrent w.r.t.~the connection $\mathfrak{D}$. As the torsion is totally 
anti-symmetric, whence the connection is metric,
the arguments of ref.~\cite{Gibbons:2007zu} can be straightforwardly generalised to see 
that the holonomy of $\mathfrak{D}$ is
contained in $Sim(3)$.
\par
We shall focus on the $\mathcal{B}=0$ case due to its special properties.

\subsection{${{\cal{B}}}=0$}
For ${{\cal{B}}}=0$, the general solution \eqref{solution} reduces to 
\begin{equation}
ds^2 = 2 du \bigg( dv +\left(H- {\chi^2 \over 8} v^2\right) du + \phi
\bigg) - ds_{GT}^2  \ , \qquad 
F = {\chi \over 4} du \wedge dv \  , \label{solutionB0}
\end{equation}
where $GT$ is the round $S^3$ with Ricci scalar $R_3 = 9
\chi^2/8$, $H$ is a harmonic function on $GT$ which may also depend on $u$ (but
not on $v$) and $\phi$ is a $u$-dependent 1-form on $GT$ (which does not depend on $v$)
satisfying 
\begin{equation}
{\hat{d}} \phi = {\sqrt{3} \chi \over 2} \star_3 \phi \ . \label{constraint1}
\end{equation}
${\hat{d}}$ denotes the exterior derivative restricted to
hypersurfaces of constant $u$, and $\star_3$ denotes the Hodge dual
on $GT$. This family of backgrounds has $Sim$ holonomy and constant scalar curvature invariants, as we shall now describe.

\subsubsection{$Sim$ holonomy}
If ${{\cal{B}}}=0$, $N$ is a {\it recurrent} null vector field; in particular we find that
\begin{equation}
\nabla_\mu N^\nu = -{1 \over 4} \chi^2 v N_\mu N^\nu \ . \label{recurrence}
\end{equation}
The recurrence relation \eqref{recurrence} is enough to show that it has holonomy $Sim(3)$ \cite{Gibbons:2007zu}. The Similitude group $Sim(n-2)$ is an $(n^2-3n+4)/2$-dimensional subgroup of the Lorentz group $SO(n-1,1)$, which is isomorphic to the Euclidean group $E(n-2)$ augmented by homotheties (or similarity transformations; hence its name). The $Sim$ group leaves invariant a null direction. Since this is the maximal proper sub-group of the Lorentz group, it is the largest holonomy group one can have for geometries with reduced holonomy. Supersymmetric geometries are expected to have reduced holonomy groups, since there are (super-)covariantly constant spinors. In the De Sitter case we are indeed finding the minimal possible (yet non-trivial) holonomy reduction. In the Minkowski and AdS theories, on the other hand, the holonomy reduction can be larger; for a generic Brinkmann wave the holonomy is just $E(n-2)$. For more details about the $Sim$ groups and geometries with $Sim$ holonomies see  \cite{Gibbons:2007zu}.

\subsubsection{Curvature and Scalar Curvature Invariants}
Let us consider the structure of the Riemann tensor. A computation shows that 
\begin{eqnarray}
\label{riem1}
R_{\mu \nu \lambda {{\tau}}u} =(R^0)_{\mu \nu \lambda {{\tau}}u}+4 N_{[\mu} \theta_{\nu][\lambda} N_{{{\tau}}u]} 
+ N_\mu \psi_{\nu \lambda {{\tau}}u}-N_\nu \psi_{\mu \lambda {{\tau}}u} + N_\lambda \psi_{{{\tau}}u \mu \nu}
-N_{{\tau}}u \psi_{\lambda \mu \nu} \ , \nonumber \\
\end{eqnarray}
where
\begin{eqnarray}
\psi_{\mu \nu \lambda} &=& {2 \over 3} \nabla_\mu \nabla_{[\nu} \phi_{\lambda]}+{1 \over 3} \nabla_\nu
\nabla_{[\mu}\phi_{\lambda]}-{1 \over 3} \nabla_\lambda \nabla_{[\mu} \phi_{\nu]} \ , 
\end{eqnarray}
\begin{equation}
\theta_{\mu \nu} = \nabla_\mu \nabla_\nu H +{1 \over 4} \chi^2 v \nabla_{(\mu} \phi_{\nu)}+{1 \over 4}
(d \phi)_{\mu \lambda} (d \phi)_\nu{}^\lambda \ , 
\end{equation}
and $(R^0)_{\mu \nu \lambda {{\tau}}u}$ denotes the Riemann tensor of $g^0$, i.e. $dS_2 \times S^3$ which is obtained
by setting $\phi=0, H=0$ in the above solution.
Note in particular that
\begin{equation}
\label{contrn}
N^\mu \psi_{\mu \nu \lambda}=0 \ , \qquad N^\mu \theta_{\mu \nu}=0 \ .
\end{equation}
From \eqref{riem1} it follows straighforwardly that 
\begin{equation}
R_{\nu{{\tau}}u}=(R^0)_{\mu\nu\lambda{{\tau}}u}g^{\mu\lambda}+\theta^\mu_\mu N_\nu N_{{\tau}}u-2\psi^\mu_{\ \mu(\nu}N_{{{\tau}}u)} \ ; \end{equation}
noting that the inverse metric $g^{\mu\nu}$ has $g^{uu}=0$ it follows that 
\begin{equation}
R_{\mu\nu}N^\mu N^\nu=0 \ , \end{equation}
which is expected for Kundt geometries \cite{exact}. From the expression of the Ricci tensor it follows that 
\begin{equation} R=(R^0)_{\mu\nu\lambda{{\tau}}u}g^{\mu\lambda}g^{\nu{{\tau}}u}=(R^0)_{\mu\nu\lambda{{\tau}}u}(g^0)^{\mu\lambda}(g^0)^{\nu{{\tau}}u}=R^0 , \label{ricciscalar} \end{equation}
where $R^0$ is the Ricci scalar of $g^0$. The middle equality follows from an analysis of the non-trivial components of $g^{\mu\nu}$ and $(R^0)_{\mu \nu \lambda {{\tau}}u}$. The latter is the direct sum of the curvature tensors for $dS_2$ and for the 3-sphere; the full inverse metric and the one for $g^0$ obey  
\begin{equation} \label{g0} g^{uv}=(g^0)^{uv} \ , \qquad  g^{ij}=(g^0)^{ij}\ , \end{equation} 
where $x^i$ are the coordinates on $S^3$.

Actually, the geometry \eqref{solutionB0} has an interesting property which generalises \eqref{ricciscalar}: all scalar invariants constructed solely from the Riemann curvature and the metric (i.e. without covariant derivatives) are constant and equal to the analogous scalar invariant for $g^0$. If we denote such scalar invariant, of degree $p$, by $S^{(p)}$, then the statement is:
\begin{equation}
{S}^{(p)} \left[(R^0)_{\mu \nu \alpha \beta}, g_{{{\tau}}u \sigma}\right] = {S}^{(p)} \left[(R^0)_{\mu \nu \alpha \beta}, (g^0)_{{{\tau}}u \sigma}\right] \ .
\end{equation}

Let us prove this. Consider any scalar invariant of degree $p$. Note that from inspection of 
({\ref{riem1}}), and using ({\ref{contrn}}), it follows that such a scalar invariant can be written
schematically as

\begin{equation}
{S}^{(p)} =c_p (R^0)^p + c_{p-1} (R^0)^{p-1} + \dots + c_1 R^0 + c_0 \ ,
\end{equation}
where $c_{p-k}$ is of degree $k$ in $\theta$, $\psi$. The proof follows in three steps:

\begin{enumerate}
\item[i)] First note that $c_0=0$; this follows directly from ({\ref{contrn}}).
\item[ii)] Secondly, note that for $1 \leq k <p$, 
$c_{p-k} (R^0)^{p-k}$ must vanish. This is because if there is any contraction
of $N$ with $R^0$, the only corresponding component of $R^0$ entering into such a 
contraction is $(R^0)_{uvuv}$ (corresponding to the $dS_2$ Riemann tensor).
Observe that in the $dS_2$ Riemann tensor, there is a pairing between $u$ and $v$ indices;
the $N$ contraction eliminates one of the $v$ indices, leaving one unpaired $u$ index; this must
contract with a tensor containing one free contravariant $u$ index. Such an object cannot be constructed
from $\theta$ or $\psi$ because $\theta^u{}_\alpha=0$, $\psi^u{}_{\alpha \beta}= \psi_\alpha{}^u{}_\beta=0$,
where indices are raised with respect to the metric given in \eqref{solutionB0}. Hence, all
such contractions must vanish.
\item[iii)] Finally, having eliminated these terms, it follows that the curvature invariant is
constructed entirely from $R^0$, but with indices raised using the metric given in
\eqref{solutionB0}. However, \eqref{g0} shows that the $\phi$ and $H$ terms in this metric
do not give any contribution to this expression; the only components of the inverse metric
entering into ${S}^{(p)}$ are $g^{uv}$, and $g^{ij}$ which are identical to the corresponding 
inverse metric components in $dS_2 \times S^3$ (i.e. there is no dependence on $H$ or $\phi$
in these components).  Hence the curvature invariant corresponds to a scalar curvature invariant of $dS_2 \times S^3$ which is constant, since it is a direct product of constant curvature spaces.
\end{enumerate}

The solution \eqref{solutionB0} seems to have further interesting properties concerning scalar invariants that contain covariant derivatives. Indeed, computing some examples appears to indicate that they all vanish. It appears, therefore, to be a five dimensional example of the four dimensional I-symmetric spacetimes studied in \cite{Coley:2009tx}.\footnote{We thank S. Hervik for bringing this reference to our attention.} The reader interested in spacetimes with vanishing or constant scalar invariants is referred to \cite{Coley:2008th} and references therein.

\subsubsection{Special Cases}
\label{nariai}

It has already been observed that, when ${\cal{B}}=0$, $\phi=0$, $H=0$, the geometry is just $dS_2 \times S^3$.  In four dimensional General Relativity, the direct product of a two dimensional De Sitter space and a two dimensional sphere, together with a flux proportional to the volume form of either the De Sitter space or the sphere,  is a very simple solution to Einstein-Maxwell theory with a positive cosmological constant, called the Nariai solution \cite{Nariai}. It is straightforward to generalise this solution to higher dimensional Einstein-Maxwell-$\Lambda$ theory \cite{Cardoso:2004uz}. Since for this solution the Chern-Simons term is irrelevant, it also arises in the minimal De Sitter supergravity we are considering herein.

Solutions with
$\phi=0$, $H \neq 0$ correspond to plane fronted gravitational waves  (rather than pp-waves, since the null vector is not covariantly constant) propagating on $dS_2 \times S^3$. 

In order to construct examples of solutions with $\phi \neq 0$, it is convenient
to write the metric on $GT$ as
\begin{equation}
ds_{GT}^2 = {4 \over 3 \chi^2 } \big[ (\sigma^1_L)^2+ (\sigma^2_L)^2+ (\sigma^3_L)^2 \big] \ , 
\end{equation}
where $\sigma^i_L$ are the standard left invariant 1-forms on $SU(2)$ satisfying 
\begin{equation}
d \sigma^i_L = {1 \over 2}\epsilon^{ijk} \sigma^j_L \wedge \sigma^k_L \ , 
\end{equation}
then a solution to ({\ref{constraint1}}) is obtained by setting
\begin{equation}
\phi = \xi_i(u) \sigma^i_L \ , 
\end{equation}
where $\xi_i$ depend only on $u$. This gives rise to a squashing of the five dimensional Nariai solution, on top of which plane fronted gravitational waves may propagate, if we take a non trivial $H$.

\subsection{An example with ${{\cal{B}}}\neq 0$}
An example with ${{\cal{B}}} \neq 0$ can be constructed by taking the Gauduchon-Tod space to be the Berger sphere. As observed in \cite[prop. 6]{tod}, the Berger sphere is the only
compact GT-space that is not an Einstein space and has non-vanishing Weyl-scalar;\footnote{
  The Weyl scalar is constrained, by pseudo supersymmetry, to be $\mathtt{W}=-9\chi^{2}/8$,
  which is non-vanishing.} Using the explicit GT-structure of the Berger sphere, eqs.~(\ref{eq:W6a}) given in appendix C, we find a solution, which is another squashing of the Nariai cosmos. Explicitly the solution reads
\begin{eqnarray}
  \label{eq:NariaiS1}
  ds^{2}  = 2du\left( dv \ -\ \frac{\chi^{2}}{8}v^{2}\ du
                    \ +\ v\sin \mu\cos\mu \ \sigma_L^3 
                 \right) 
         \ -\ \frac{4\cos^{2}\mu }{3\chi^{2}}\ ds^2_{\rm Berger} \; , \nonumber \\
  A       = -\frac{\chi}{4}\ v\ du
         \ +\ \frac{\sin \mu\cos\mu }{\chi}\ \sigma_L^3 \; . \qquad \qquad \qquad \qquad
\end{eqnarray}
For $\mu=0$ we recover the Nariai solution of section \ref{nariai}.

\section{Final Remarks}
In this paper the most general null solution of $D=5$ minimal De Sitter supergravity admitting (pseudo-)Killing spinors was computed. The solutions we found preserve at least half of the (pseudo)-Killing spinors and can be described as a particular type of (higher dimensional generalisation of) Kundt geometry. Unlike the null solutions of minimal ungauged and gauged supergravity, the null vector $N$, which is geodesic, twist-free, shear-free and expansion-free is not Killing. This is analogous to what happens in the timelike case. The timelike vector field built from Killing spinors is Killing in the ungauged and gauged supergravity theories but not in the De Sitter theory \cite{guthktd5}. 

The null vector $N$ can, however, have a special property: it may become recurrent. This means that the reduction of the holonomy group of the geometries is minimal. The geometries then have $Sim(3)$ holonomy, the maximal proper sub-group of the five dimensional Lorentz group. The results of \cite{Meessen:2009ma,Gutowski:2009vb} show that the null solutions of $D=4$ minimal De Sitter supergravity admitting (pseudo-)Killing spinors have holonomy $Sim(2)$. It would be interesting to know if (pseudo-)supersymmetric null solutions in De Sitter supergravity in all dimensions admit, at least for a subset of the solutions, $Sim$ holonomy. 

In ref.~\cite{Gibbons:2007zu}, Gibbons and Pope showed that the dimensional reduction of  a space
of $Sim$ holonomy along a spacelike direction in the lightcone gives rise to time-dependent Kaluza-Klein black holes. The general solution found here can, as well, be dimensionally reduced to four 
dimensions, leading to backgrounds that fit naturally in the general class of solutions found in
refs.~\cite{Meessen:2009ma,Gutowski:2009vb}. As ref.~\cite{Gibbons:2007zu} focusses on 
spaces with $Sim$ holonomy, the dimensionally reduced solutions miss the characteristic 
time-dependence associated to a non-vanishing ${{\cal{B}}}$. Thus, the dimensional reduction of the Kundt metrics found here leads to time-dependent KK black-holes with a more general time dependence than the ones considered in \cite{Gibbons:2007zu}.

\acknowledgments
We thank M.~Dunajski, A.~Chatterjee and T.~Ort\'{\i}n for fruitful discussions. Jai Grover thanks the Cambridge Commonwealth
Trusts for financial support. Jan Gutowski is supported by the EPSRC grant, EP/F069774/1.
C.H. is supported by a ``Ci\^encia 2007" research contract from ``Funda\c c\~ao para a Ci\^encia e a Tecnologia". P.M. is supported by the Spanish Ministry of Science through the project FPA2006-00783.
A.P.L. is supported by a C.S.I.C.~scholarship JAEPre-07-00176. This work as been further supported by the Comunidad de Madrid grant HEPHACOS P-ESP-00346, by the EU Research Training Network \textit{Constituents, Fundamental Forces and Symmetries of the Universe} MRTN-CT-2004-005104, and by the Spanish Consolider-Ingenio 2010 program CPAN CSD2007-00042.

\appendix

\section{The Linear System}

The gravitino equation acting on $\epsilon $ in the $+$ direction
gives
\begin{equation}
{\chi }A_{+} = -{i\over\sqrt{3}}F_{+2} -(\omega_{+,+-} +
\omega_{+,1\bar{1}})\  ,  \nonumber
\end{equation}
\begin{equation}
-{i\over\sqrt{3}}F_{+\bar{1}}+\omega_{+,\bar{1}2}=0\ ,  \nonumber
\end{equation}
\begin{equation}
\omega _{+,+2}=0\ ,  \nonumber
\end{equation}
\begin{equation}
\omega _{+,+1}=0\ .  
\end{equation}
In the $-$ direction
\begin{equation}
{\chi }A_{-} = \sqrt{3}iF_{-2} -(\omega_{-,+-} +
\omega_{-,1\bar{1}})\ , \nonumber
\end{equation}
\begin{equation}
-\sqrt{3}iF_{-\bar{1}}+\omega_{-,\bar{1}2}=0 \ ,  \nonumber
\end{equation}
\begin{equation}
{2\over\sqrt{3}}F_{+-} - i\omega_{-,+2}
-{1\over\sqrt{3}}F_{1\bar{1}}+{\chi\over2\sqrt{3}}=0\ ,  \nonumber
\end{equation}
\begin{equation}
i\omega _{-,+\bar{1}}+{1\over\sqrt{3}}F_{\bar{1}2}=0\ .  
\end{equation}
In the $1$ direction
\begin{equation}
{\chi }A_{1} = \sqrt{3}iF_{12}-(\omega_{1,+-} +
\omega_{1,1\bar{1}})\ ,  \nonumber
\end{equation}
\begin{equation}
-{2i\over\sqrt{3}}F_{1\bar{1}}+\omega_{1,\bar{1}2}+{i\over\sqrt{3}}F_{+-}-{i\chi\over2\sqrt{3}}=0\
,  \nonumber
\end{equation}
\begin{equation}
\sqrt{3}F_{+1}-i\omega_{1,+2}=0\ ,  \nonumber
\end{equation}
\begin{equation}
i\omega_{1,+\bar{1}}+{1\over\sqrt{3}}F_{+2}=0\ .  
\end{equation}
In the $\bar{1}$ direction
\begin{equation}
{\chi }A_{\bar{1}} =
{i\over\sqrt{3}}F_{\bar{1}2}-(\omega_{\bar{1},+-}+\omega_{\bar{1},1\bar{1}})=0\
,  \nonumber
\end{equation}
\begin{equation}
\omega_{\bar{1},\bar{1}2}=\omega_{\bar{1},+\bar{1}}=0\ ,  \nonumber
\end{equation}
\begin{equation}
{1\over\sqrt{3}}F_{+\bar{1}}- i\omega_{\bar{1},+2}=0\ ,  
\end{equation}
In the $2$ direction
\begin{equation}
{\chi }A_{2} = -{i\over\sqrt{3}}(F_{+-} + F_{1\bar{1}})
-(\omega_{2,+-} + \omega_{2,1\bar{1}})+{i\chi\over2\sqrt{3}}\ ,  \nonumber
\end{equation}
\begin{equation}
{2i\over\sqrt{3}}F_{\bar{1}2}+\omega _{2,\bar{1}2}=0\ ,  \nonumber
\end{equation}
\begin{equation}
{2\over\sqrt{3}}F_{+2}-i\omega _{2,+2}=0\ ,  \nonumber
\end{equation}
\begin{equation}
i\omega _{2,+\bar{1}}-{1\over\sqrt{3}}F_{+\bar{1}}=0\ .  
\end{equation}

\section{$Spin(4,1)$ Gauge Transformations}

The most general $Spin(4,1)$ gauge transformation preserving the
Killing spinor $\epsilon = 1+e_1$ is generated by
\begin{equation}
T_1 = \gamma_{01}+\gamma_{13}, \quad T_2 = \gamma_{02}+\gamma_{23},
\quad T_3 = \gamma_{04} - \gamma_{34} \ ,  
\end{equation}
which satisfy
\begin{equation}
T_i (1+e_1) =0 \ .
\end{equation}

The induced effect of the gauge transformation $e^{xT_1+yT_2+zT_3}$
(for $x,y,z \in {\mathbb{R}}$)  on the vielbein is given by
\begin{eqnarray}\label{gaugeframe}
{\mathbf{e}}^+ &\rightarrow& {\mathbf{e}}^+ + \mu {\mathbf{e}}^- + {{\tau}}u_{m}{\mathbf{e}}^{m} \ ,\nonumber \\
{\mathbf{e}}^- &\rightarrow& {\mathbf{e}}^- \ ,\nonumber \\ {\mathbf{e}}^{m} &\rightarrow& {\mathbf{e}}^{m}+
\sigma^{m}{\mathbf{e}}^- \ .  
\end{eqnarray}
where here the Latin index $m = 1, \bar{1}, 2$, and
\begin{eqnarray}
 \sigma^1 = y + iz \ , \quad
 \sigma^{\bar{1}} = y-iz \ , \quad
 \sigma^2 = \sqrt{2}x \ ,\quad
{{\tau}}u_{m} = \delta_{mn}\sigma^{n} \ , \quad
\mu={1\over2}\delta_{mn}\sigma^{m}\sigma^{n}\ .   \nonumber \\
\end{eqnarray}

Such transformations will leave the metric invariant, and can be
used to set
\begin{eqnarray}
(\mathcal{L}_{N}{\mathbf{e}}^{m})_- &\rightarrow&
(\mathcal{L}_{N}{\mathbf{e}}^{m})_- + (\mathcal{L}_{N}\sigma^{m})_- +
\sigma^{m}(\mathcal{L}_{N}{\mathbf{e}}^-)_- \nonumber \\ &=&(d{\mathbf{e}}^{m})_{+-} +
\sigma^{m}(d{\mathbf{e}}^-)_{+-} +
\partial_+\sigma^{m} \ ,  
\end{eqnarray}
which can be locally made to vanish for a suitable choice of
$\sigma$. 

Further simplification can be made by considering the
combined ${\mathbb{R}} \times Spin(4,1)$ gauge transformation
$e^{-h} e^{h \Gamma_{+-}}$ for $h \in {\mathbb{R}}$,
which also leaves the Killing spinor $\epsilon=1+e_1$ invariant.
Under this transformation, the gauge potential transforms as
\begin{equation}
A \rightarrow A - {2 \over \chi} dh \ .  
\end{equation}
This allows us to set $A_+ = -{1\over\chi}\omega_{+,+-} = 0$ without
loss of generality. Hence we also find that
\begin{equation}
(\mathcal{L}_{N}{\mathbf{e}}^-)_- = 0 \ .  
\end{equation}

\section{Gauduchon-Tod spaces}
\label{sec:EWspaces}

A Weyl manifold is a manifold $\mathcal{M}$ of
dimension $n$ together with 
a conformal class $[g]$ of metrics on $\mathcal{M}$ and a torsionless
connection $\mathtt{D}$, which preserves the conformal class, {\em
  i.e.\/}
\begin{equation}
  \label{eq:W1}
     \mathtt{D}\ g \; =\; 2\theta \otimes g \; , 
\end{equation}
for a chosen representative $g\in [g]$. 
Using the above definition, we can express the connection $\mathtt{D}_{X}Y$
as 
\begin{equation}
  \label{eq:W2}
      \mathtt{D}_{a}Y_{b} \; =\; \nabla^{g}_{a}Y_{b} \ +\
      \gamma_{ab}{}^{c}\ Y_{c} \hspace{.5cm}\mbox{with}\hspace{.5cm}
      \gamma_{ab}{}^{c} \ =\ g_{a}{}^{c}\theta_{b} \ +\ g_{b}{}^{c}\theta_{a} \ -\ g_{ab}\theta^{c}\; ,
\end{equation}
where $\nabla^{g}$ is the Levi-Civit\`a connection for the chosen
$g\in [g]$. We define the curvature of this connection through
$\left[ \mathtt{D}_{a},\mathtt{D}_{b}\right] Y_{c} =
-\mathtt{W}_{abc}{}^{d}Y_{d}$, using which we define
the associated Ricci curvature as $\mathtt{W}_{ab} \equiv \mathtt{W}_{acb}{}^{c}$.
The Ricci tensor is not symmetric
and we have
\begin{eqnarray}
  \label{eq:W3a}
     \mathtt{W}_{[ab]}  = -\textstyle{n\over 2}\
     F_{ab}\ , \ \ \mbox{where}\ \ \  F\equiv d\theta \; , 
      \qquad \qquad \qquad \qquad \\
       \nonumber  \\
     \label{eq:W3b}
     \mathtt{W}_{(ab)}  = \mathtt{R}(g)_{ab}
           \ -\ (n-2) \nabla_{(a}\theta_{b)}
           \ -\ (n-2)\ \theta_{a}\theta_{b}
           \ -\ g_{ab}\left[
                   \nabla_{a}\theta^{a} 
                   \ -\ (n-2)\ \theta_{a}\theta^{a}
                 \right] \; \ . \nonumber \\
\end{eqnarray}
The Weyl-scalar is defined as $\mathtt{W}\equiv
\mathtt{W}_{a}{}^{a}$, which explicitly reads
\begin{equation}
  \label{eq:W4}
      \mathtt{W} \; =\; \mathtt{R}(g) \ -\ 2(n-1)\ \nabla_{a}\theta^{a}
               \, +\, (n-1)(n-2)\
               \theta_{a}\theta^{a} \; .
\end{equation}
The 1-form $\theta$ acts as gauge field
gauging an $\mathbb{R}$-symmetry, which is the reason why we
have a conformal class of metrics on $\mathcal{M}$;
in fact under a transformation $g\rightarrow e^{2w}\ g$ we have that
$\theta\rightarrow \theta +dw$ and $\mathtt{W}\rightarrow
e^{-2w}\mathtt{W}$,
whereas $\mathtt{W}_{abc}{}^{d}$ and $\mathtt{W}_{ab}$ are conformally
invariant.
\par
A Weyl manifold is said to be {\em Einstein-Weyl} if the curvatures satisfy
\begin{equation}
  \label{eq:W5}
  \mathtt{W}_{(ab)} \; =\; \frac{1}{n}\ g_{ab}\ \mathtt{W} \; .
\end{equation}
\par
A metric $g$ in the conformal class $[g]$ is said to be {\em standard} or {\em Gauduchon} if
it is such that 
\begin{equation}
  \label{eq:W4a}
  d\star\theta \ =\ 0 \hspace{.4cm}\mbox{or equivalently}\hspace{.4cm}
  \nabla_{a}\theta^{a}\ =\ 0\; ,
\end{equation}
where the $\star$ is taken w.r.t.~the chosen metric $g$.
Gauduchon \cite{art:gauduchon1984a} proved the existence of a standard metric compact EW manifold, and 
Tod \cite{art:Tod1992} proved that on compact EW manifolds this implies that $\theta^{\flat}$
is a Killing vector of the standard metric $g$.
\par
In ref.~\cite{tod}, Gauduchon {\&} Tod studied the structure of 4-dimensional hyper-Hermitian
Riemannian spaces admitting a tri-holomorphic Killing vector, {\em i.e.} Killing vectors that are compatible with
the 3 complex structures on the hyperHermitian space. A result of that study is that the 3-dimensional base-space 
is determined by a Dreibein, or orthonormal frame, $E^{x}$, a 1-form $\theta$ and a real function 
$\kappa$ that must satisfy
\begin{equation}
  \label{eq:GTa}
  dE^{x} \; =\; \theta\wedge E^{x} \ -\ \kappa\ \star E^{x} \; ,
\end{equation}
where $\star$ is to be taken w.r.t.~the Riemannian metric constructed out of the Dreibein.
The underlying geometry imposed by the above equation is that of a specific type of
3-dimensional EW-spaces, called hyperCR or Gauduchon-Tod spaces.\footnote{
   Observe that the Jones-Tod construction \cite{art:Tod1985} implies that the 3-dimensional
   space orthogonal to a generic Killing vector on a 4-dimensional hyperHermitian space
   is always Einstein-Weyl.
}
The extra restriction to be imposed on the EW-space, which are equivalent to imposing (\ref{eq:GTa}),
are\footnote{
  The sign difference between eq.~(\ref{eq:GTb}) and eq.~(S) in \cite[prop.~5]{tod}
  is due to a differing definition of the Riemann tensor.
}
\begin{eqnarray}
  \label{eq:GTb}
  \mathtt{W} & =& -\frac{3}{2}\ \kappa^{2}\; ,\\
  \label{eq:GTc} 
  \star d\theta & =& d\kappa\ +\ \kappa\ \theta \; . 
\end{eqnarray}
Comparing this last expression to the ones obtained in the main text, we see that $\kappa =-\sqrt{3}\chi/2$
and $\theta = \chi {{\cal{B}}}$ (note also that the operator $\star$ corresponds to $-\star_3$ in the main text).
\par
The standard example \cite{tod} of a GT-space is the {\em Berger sphere}
\begin{eqnarray}
  \label{eq:W6a}
  ds^{2}_{\rm Berger}  & =& d\phi^{2}\ +\ \sin^{2}\phi d\varphi^{2}\ +\
              \cos^{2}\mu \left[ d\psi\ +\ \cos \phi d\varphi \right]^{2}=(\sigma_L^1)^2+(\sigma_L^2)^2+\cos^2\mu (\sigma_L^3)^2
        \; ,\nonumber \\
  \theta & =& \sin \mu \cos \mu \ \left[ d\psi\ +\ \cos \phi d\varphi \right]=\sin\mu\cos\mu \ \sigma_L^3\; ,
\end{eqnarray}
which is a squashed $S^{3}$ or an $\mathrm{SU}(2)$ group manifold with a $\mathrm{U}(1)$-invariant
metric. One can easily see that the metric is Gauduchon-Tod with 
$\kappa = \cos\mu $: this means that in order to use it in the 5-dimensional solutions
it needs to be rescaled by a constant.
\par
Another class of GT-spaces, albeit not in the Gauduchon-gauge, was found by Calderbank and Tod \cite{calderbank}
and reads
\begin{eqnarray}
  \label{eq:Cald1}
  ds^{2} & =& dx^{2} \, +\, 4\left| x+h\right|^{2}\ \frac{dzd\bar{z}}{(1+|z|^{2})^{2}} \; ,\\
  \label{eq:Cald2}
  \theta & =& 2\mathrm{Re}\left(\textstyle{1\over x+h}\right)\ dx\; ,\\
  \label{eq:Cald3}
  \kappa & =& 2\mathrm{Im}\left(\textstyle{1\over x+h}\right) \; ,
\end{eqnarray}
where $h$ is an arbitrary holomorphic function $h=h(z)$.
As $\kappa$ is not constant, we must rescale the metric in order to use it to construct a 5-dimensional solution.
Observe that the choice $h=-\bar{h}$ results in the 3-sphere and that the 
choice $h=\bar{h}$ leads to the flat metric on $\mathbb{R}^{3}$ with $\kappa =0$ and cannot be used to generate
5-dimensional solutions.

\end{document}